# All-chalcogenide Raman-parametric Laser, Wavelength Converter and Amplifier in a Single Microwire

Raja Ahmad[*] and Martin Rochette

Department of Electrical and Computer Engineering, McGill University, Montreal (QC), Canada, H3A 2A7.

**Compact, power efficient and fiber-compatible lasers, wavelength converters and amplifiers are vital ingredients for the future fiber-optic systems and networks. Nonlinear optical effects, like Raman scattering and parametric four-wave mixing, offer a way to realize such devices. Here we use a single chalcogenide microwire to realize a device that provides the functions of a Stokes Raman-parametric laser, a four-wave mixing anti-Stokes wavelength converter, and an ultra-broadband Stokes/anti-Stokes Raman amplifier or supercontinuum generator. The device operation relies on ultrahigh Raman and Kerr gain (upto five orders of magnitude larger than in silica fibers), precisely engineered chromatic dispersion and high photosensitivity of the chalcogenide microwire. The Raman-parametric laser operates at a record low threshold average (peak) pump power of 52 µW (207 mW) and a slope efficiency of >2%. A powerful anti-Stokes signal is generated via the nonlinear four-wave mixing process. As amplifier or the broadband source, the device covers a wavelength (frequency) range of >330 nm (47 THz) when pumped at a wavelength of 1550 nm. Owing to the underlying principle of operation of the device being the nonlinear optical processes, the device is anticipated to operate over the entire transmission window of the chalcogenide glass ($\lambda \sim 1\text{-}10$ µm).**

The future of fiber-optic systems lies in the development of all-fiber photonic devices that are compact, power efficient, and can provide novel functionalities. Optical microwires (μ-wires) are fiber waveguides with cross-section diameters in the order of the operating wavelength. The μ-wires are attractive owing to their negligible coupling and propagation losses [1], orders of magnitude enhancement in waveguide nonlinear coefficient [2], controllable amount of chromatic dispersion [3] and evanescent optical field propagation [4]. These advantages have driven researchers to use μ-wires for studying novel low field optical phenomena [5], as well as for wide-ranged application in the fields of nonlinear optics [6], sensing [7] and plasmonics [8], to name a few.

Optical sources, both coherent and incoherent, are among the basic components of any fiber-optic system. The former are narrowband and are usually lasers or their wavelength converted counterparts (via nonlinear wave-mixing process), while the latter are broadband, rely on spontaneous emission processes and can act as the amplifiers or as broadband light sources. Recently, lasing has been reported in μ-wires made from rare-earth or dye doped silica glass [9], but these devices required pumping via evanescent coupling to a μ-wire coiled in a resonator geometry. The evanescent coupling is unfortunately an obstacle to the practical use of such lasers in fiber systems, where the large coupling and/or transmission losses of such devices cannot be endured. Moreover, the gain spectrum following from discrete atomic transitions in doped silica limits the wavelength tunability of these devices.

In contrast with this, the nonlinear optical effects manifest at arbitrary wavelengths. This offers a way to circumvent the issue of wavelength tunability. The $As_2Se_3$ chalcogenide glass fiber exhibits nonlinear Kerr and Raman gain coefficients that are up to 3 orders of magnitude larger than that of silica fibers [10]. The nonlinear gain, defined as the ratio of output power to the input power, is further enhanced in a μ-wire geometry, where the optical field intensity is enhanced by more than 100× due to the correspondingly reduced mode area. Therefore, $As_2Se_3$ μ-wires are excellent candidates for the development of compact and efficient sources based on nonlinear optical effects. The $As_2Se_3$ glass also carries the advantage of a large photosensitivity [11], enabling the direct inscription of Bragg grating (BG) reflectors within the μ-wire structure to complete the laser cavity [12]. Of great importance also carried by $As_2Se_3$ is an ultra-wide transmission window that allows the useful operation spectrum over the wavelength range of $\lambda \sim 1$–10 μm [13].

Recently, we made the first and only reports of lasers based on the nonlinear gain in μ-wires [14], showing the use of the Raman gain in combination with an external resonant cavity to generate the laser effect. Unfortunately, it appears that the round-trip cavity losses could be reduced by using more efficient resonant cavities, the use of free-space components could be replaced into a complete integrated solution, and that silica components could be replaced to broaden the operation window further towards the mid infrared.

In this Letter, we report a distributed Bragg reflector (DBR) type $As_2Se_3$ μ-wire Raman-parametric laser, wavelength converter, and amplifier in a single chalcogenide microwire. The resonant cavity of the device is made out of two BGs directly written within the $As_2Se_3$ μ-wire, exploiting its high photosensitivity. This allows the integration of the nonlinear gain medium and the cavity mirrors in a single μ-wire, thus minimizing the cavity losses and power threshold of the laser, as well as to improve the slope efficiency. The combination of nonlinear gain and BGs in a single microwire makes the device capable of operating over the entire transmission window of $As_2Se_3$ glass (1–10 μm). The two ends of the μ-wire laser are adiabatically mode-matched to single mode silica fiber and thus make the device compatible with optical fibers and waveguides [15]. The DBR-laser operates at the Stokes Raman wavelength, where the BGs form a resonant cavity. The chromatic dispersion in the μ-wire is carefully controlled to also achieve four-wave mixing (FWM) based wavelength conversion of the Raman laser signal into the anti-Stokes wavelength via interaction with the Raman input pump. The sole effect of the µ-wire as a travelling wave amplifier and supercontinuum generator is also investigated. In this context, multiple copies of Raman gain spectrum are observed at both anti-Stokes (5 orders) and Stokes wavelengths (≥ 2 orders) via FWM. This multi-frequency Raman-parametric spectrum spans over a wavelength (frequency) range of more than 330 nm (47 THz). In addition to providing the functionalities of a laser and wavelength converter, the resulting device thus acts as an ultra-broadband multi-order, Stokes/anti-Stokes Raman amplifier and/or a supercontinuum source.

*Experiment and results:* Fig. 1 shows the setup for the operation of the µ-wire Raman laser. The pump source is a continuous-wave (CW), external cavity laser, tunable within the telecommunications C-band. The pump laser output is passed through a pulse-carving stage with an extinction ratio of ≥36 dB. The pulse carving stage consists of two cascaded Mach-Zehnder modulators fed with identical electrical data inputs from a pulse pattern generator. The power level of the prepared pulses is adjusted to the required levels by using a series of two erbium-doped fiber amplifiers (EDFAs) and two bandpass filters with <0.4 nm passband, followed by an optical attenuator. The figure also includes a schematic of the µ-wire laser device, which is an $As_2Se_3$ µ-wire with BGs inscribed at its two ends. The µ-wires are fabricated using a modified flame brush technique [15] and the BGs are inscribed using a He-Ne laser ($\lambda$ = 633 nm) based transverse holographic setup assisted with a glass prism [12]. After being permanently bonded to standard single mode silica fibers via UV-curing and being inscribed with BGs, the µ-wire laser device has a total loss of 5.9 dB. The diameters of the µ-wires are carefully selected to control the amount of chromatic dispersion [16], which allows the simultaneous operation of the device as a Raman-parametric laser and a FWM wavelength converter. For this purpose, the diameter of µ-wires used in the current experiment is typically around 1.0 µm.

In a first series of experiment, the μ-wires are being used as travelling wave amplifiers without BGs, and the impact of the μ-wire diameter is studied. Each μ-wire is 13 cm in length, and is pumped by pulses that are 100 ps in duration, have a repetition rate of 1.22 MHz and are spectrally centered at a wavelength of λ = 1530 nm. Fig. 2 (a) shows the resulting single-pass output spectra through a 1.0 μm diameter μ-wire as a function of increasing pump power levels. The zero-dispersion wavelength $\lambda_{ZDW}$ for the 1.0 um diameter μ-wire lies at λ = 1529 nm, as shown in the numerically calculated dispersion profile included as the figure inset. This translates to the pump laser lying very close to the zero-dispersion wavelength and experiencing anomalous dispersion ($\beta_2 < 0$), thus setting up the optimal conditions for parametric interactions [17]. As a result, up to five parametrically converted output orders are observed at the anti-Stokes wavelengths, while two such converted orders are observed at the Stokes wavelengths. The next Stokes output signals, if present, are expected to lie beyond the operating wavelength range of the OSA.

The performance of μ-wires is also tested when pumped away from $\lambda_{ZDW}$, both in anomalous and normal dispersion regions. A μ-wire of smaller diameter, with 0.95 μm, is first used and the spectra obtained at different pump power levels are shown in Fig. 2 (b). In this case, the $\lambda_{ZDW}$ is blue-shifted and the pump still lies in the anomalous dispersion wavelength region. Almost the same number of total output wavelengths are generated as with the 1.0 μm μ-wire. It is however, noted that the spectral energy now appears to be shifted towards shorter wavelengths, and the spectrum is relatively flat with respect to the case for 1.0 μm diameter μ-wire. The spectral flatness is due to the observed blue-shift of the spectral energy, which leads from the corresponding blue-shift of the $\lambda_{ZDW}$. The Raman-FWM supercontinuum-like spectrum for both 1.0 μm and 0.95 μm diameter μ-wires, spans almost the same wavelength (frequency) range of >330 nm (47 THz). The apparently comparable bandwidth can be explained from the blue-shift of energy in the case of 0.95 μm μ-wire. To repeat the experiment on a μ-wire with >1.0 μm diameter, a μ-wire with 1.02 μm is prepared. The spectra for this μ-wire as a function of the pump power level are provided in Fig. 2 (c). As expected, fewer, that is up to 3 anti-Stokes outputs are generated in this case, due to the input pump wavelength lying in the normal dispersion ($\beta_2 > 0$) region.

To realize a laser, two BGs spatially apart by 11 cm are inscribed into a μ-wire with a diameter of 1.0 μm and a length of 13 cm. The reflection maxima of the two BGs are centered at $\lambda \sim 1585$ nm (L-band) with reflection coefficients of ~90% and 60% at the input and output ends of the μ-wire, respectively. The pump laser is centered at $\lambda \sim 1532$ nm, and the pulse duration and repetition rate are adjusted to 64 ns and ~3.8 kHz, respectively. The spatial length of the pulses propagating in the μ-wire is >7.5 meters, which allows the storage and/or amplification of the generated Raman/FWM gain in the μ-wire resonator for several round-trips (>34, in total) and leads to a quasi-continuous wave operation of the laser. As the device is pumped, Raman lasing is observed at a wavelength of $\lambda \sim 1585$ nm that is on the Stokes wavelength side of pump wavelength as shown in Fig. 3(a). The precise wavelength of operation of the Raman laser is defined by the central wavelength of the BGs. The FWM led-wavelength conversion is also visible at the anti-Stokes wavelength $\lambda \sim 1482$ nm. This spectrum represents the simultaneous conversion of a C-band pump laser to L-and S-bands frequency spectra. Fig. 3(b) plots the evolution of Stokes Raman and anti-Stokes FWM generated signals with increasing input pump power, revealing the respective slope efficiencies of 2.15% and 0.46%, and a threshold average (peak) pump power of ~52 μW (<207 mW). This represents the lowest threshold Raman laser to date in a fiber geometry [14, 18], and is also the first demonstration of a fully fiberized microwire laser of any kind. It is emphasized that the laser slope efficiency of >2% is a remarkably high value for such a compact, centimeters long μ-wire laser, operating at such low power levels.

Finally, we test the wavelength tunability of the lasing device. Fig. 4 (a) summarizes the results of the wavelength tuning experiment. By tuning the pump wavelength over a range of 8 nm ($\lambda$=1527-1535 nm), the anti-Stokes output is tuned by 14 nm ($\lambda$=1473-1487 nm), while the Stokes Raman wavelength remains fixed, bounded by the fixed wavelengths of the BGs. The laser ceases to operate when the pump laser is tuned beyond the stated wavelength range of ~8 nm, which is comparable to the Raman gain bandwidth [10]. The laser can be expected to operate so long as the Raman gain overlaps the reflection spectrum of the BGs. The asymmetric wavelength tuning between the pump laser and the anti-Stokes idler output leads from the phase matching conditions naturally satisfied during the FWM process. The slope efficiencies of the Stokes and anti-Stokes outputs are plotted in Fig. 4(b), as a function of the wavelength separation between the pump laser and the Stokes Raman laser. The slope efficiencies are maximized for a wavelength separation of ~53.5 nm, and decrease in value as the pump laser is detuned to either Stokes or anti-Stokes wavelengths.

*Discussion:* The interplay between Raman and parametric processes in nonlinear media has been studied in the past in the context of amplification and wavelength conversion, and is reported to significantly improve the net available gain value and bandwidth in a nonlinear amplifier and wavelength converter [19]. The Raman-parametric lasing however has never been reported heretofore. The device presented in this paper utilizes both Raman and parametric gains to operate simultaneously as a laser and an anti-Stokes wavelength converter. It is emphasized that the presented device is different from the reported Raman-assisted parametric amplifiers, where the Raman scattering assists in phase matching in an otherwise non-phase matched, normally dispersive media [19]. Both Raman and parametric gains at the target wavelength are present in the µ-wire from the beginning, and merely reinforce each other into simultaneous lasing and wavelength conversion.

In order to compare the results with theory, the required threshold pump power is estimated from the roundtrip cavity loss and the other parameters of the μ-wire laser, including the effective mode area and length of the μ-wire as well as the gain coefficients of the nonlinear Raman/parametric processes. The roundtrip cavity loss is estimated at ~4.7 dB and includes 2.7 dB gratings reflectivity loss and a 2 dB round-trip propagation loss. The Stokes Raman signal experiences a bidirectional gain because the spatial length of the pump pulse is much longer than the laser cavity length. The roundtrip gain therefore consists of two components: in forward propagation, the signal co-propagates with the pump and thus Raman-parametric gain acts on it, while in counter propagation the parametric gain is absent and the gain originates solely from the Raman effect. The total roundtrip gain experienced by the signal at Stokes Raman wavelength, trapped within the laser cavity, is written as,

$$G = \exp(g_{Raman-FWM} L_{eff} P_{pump} + g_{Raman} L_{eff} \frac{P_{pump}}{A_{eff}}) \qquad (1)$$

where $P_{pump}$ is the input peak pump power, $L_{eff}$ is the effective length of the μ-wire laser that is ~9.7 cm corresponding to the propagation loss coefficient $\alpha \sim 10$ dB/m, and $A_{eff}$ is the effective transverse area of the fundamental mode, and is estimated to be 0.51 μm² in the 1.0 μm diameter μ-wire. The Raman gain coefficient value for $As_2Se_3$ glass, as estimated in ref. [20], is $g_{Raman} = 2.3\times10^{-11}$ m/W. The combined Raman-parametric gain coefficient $g_{Raman-FWM}$ is defined as $g_{Raman-FWM} = 2\gamma\Re[\sqrt{K(2q-K)}]$ [19]. Here, the term γ represents the effective waveguide nonlinear coefficient estimated at ~99 W⁻¹-m⁻¹. The term $q = 1 - f + f\chi^{(3)}(-\Omega)$, where $f = 0.1$ is the fractional contribution of Raman susceptibility to the instantaneous Kerr effect in $As_2Se_3$ glass [21], Ω is the angular frequency difference between the pump and the Stokes/anti-Stokes signals, and $\chi^{(3)}(\Omega)$ is the Fourier transform of the complex Raman susceptibility function, with its value for Stokes Raman shift of ~6.8 THz [$\Omega = 2\pi\times6.8\times10^{12}$ rad/sec], calculated at -4.28$i$ for $As_2Se_3$ glass. Finally, the term $K = -\Delta k/2\gamma P_{pump} = -\beta_2 \Omega^2/2\gamma P_{pump}$ is the linear phase mismatch normalized to the nonlinear contribution to the mismatch, which is from the FWM in present case. The chromatic dispersion parameter $\beta_2$ at the pump wavelength $\lambda \sim 1532$ nm, is numerically estimated to be -5.6 ps$^2$/km. By using all the available values, the total roundtrip gain $G$ is evaluated from Eq. (1), as a function of peak pump power $P_{pump}$. The total roundtrip cavity loss of 4.7 dB estimated from the μ-wire laser parameters can be compensated with a $P_{pump} \sim 76$ mW. This value is smaller by a factor of more than 2, with respect to the pump peak power value estimated during the experiment. It is hypothesized that this mismatch leads from (1) the possible coupling of power to higher order modes in the μ-wire, and (2) an uncertainty in the value of pump peak power that is being estimated from the average pump power. The uncertainty in peak power value is due to the asymmetry in temporal profile of the 64 ns duration pulse that is caused by the gain saturation effects in EDFAs. Nevertheless, this indicates that in practice, the μ-wire laser presented here is most likely operating with a sub-100 mW threshold pump peak power, and the experimentally estimated threshold value of 207 mW, is an over-estimate.

*Conclusion:* In summary, the large Raman and parametric gain as well as the photosensitivity available in As$_2$Se$_3$ chalcogenide glass is utilized to realize a compact, low threshold, and high efficiency, microwire Raman laser and four-wave mixing wavelength converter in simultaneous operation. The generation of combined Raman-parametric ultra-broadband spectrum is also observed, covering the wavelength (frequency) range of >330 nm (47 THz). The device is fiber-compatible and is ready for immediate use in existing fiber systems. Moreover, the laser, being all-chalcogenide and based on nonlinear optical gain, can also be readily used in the mid-infrared wavelength spectrum to cater for the current high demand of such light sources.

The authors thank Coractive High-Tech for providing chalcogenide fibers. This work was financially supported by the Natural Sciences and Engineering Research Council of Canada (NSERC).

*raja.ahmad@mail.mcgill.ca

**FIGURE CAPTIONS**

**FIG. 1.** Experimental setup for the operation of chalcogenide microwire Raman laser and FWM wavelength converter. A schematic of the micro laser device is also drawn. CW laser: continuous-wave (CW) tunable laser; PC: fiber polarization controller; SMF: single-mode fiber; OSA: optical spectrum analyzer.

**FIG. 2.** Single-pass output spectra as a function of input pump power injected into the μ-wires with diameters of (a) 1.0 μm (b) 0.95 μm and (c) 1.02 μm. Numerically calculated second-order dispersion profiles of the corresponding μ-wires are included as inset in each illustration.

**FIG. 3.** (a) Output spectra showing the simultaneous operation of microwire Raman laser and FWM wavelength converter at different input pump power levels (b) Stokes and anti-Stokes slope efficiency curves.

**FIG. 4.** (a) Spectra of a microwire Raman laser-FWM wavelength converter at different pump wavelengths (b) Slope efficiencies of (Stokes) Raman signal and (anti-Stokes) FWM signals for varying wavelength separation between the input pump and the resulting Raman laser.

**FIGURES**

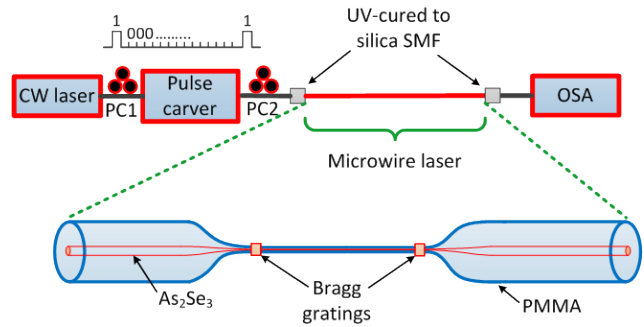

**FIG. 2.** Experimental setup for the operation of chalcogenide microwire Raman laser and FWM wavelength converter. A schematic of the micro laser device is also drawn. CW laser: continuous-wave (CW) tunable laser; PC: fiber polarization controller; SMF: single-mode fiber; OSA: optical spectrum analyzer.

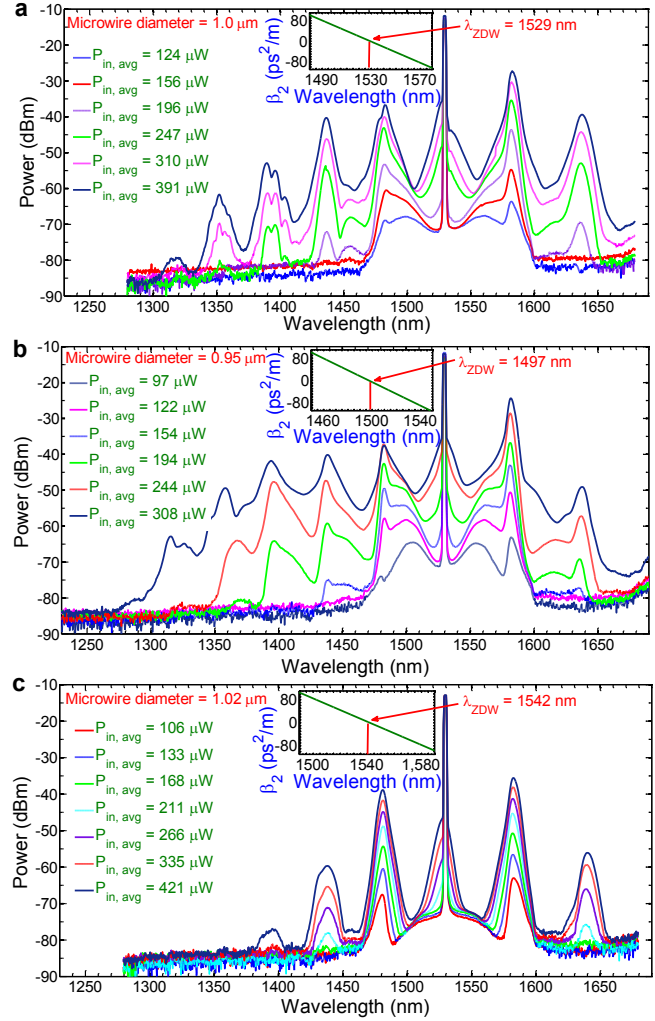

**FIG. 2.** Single-pass output spectra as a function of input pump power injected into the μ-wires with diameters of (a) 1.0 μm (b) 0.95 μm and (c) 1.02 μm. Numerically calculated second-order dispersion profiles of the corresponding μ-wires are included as inset in each illustration.

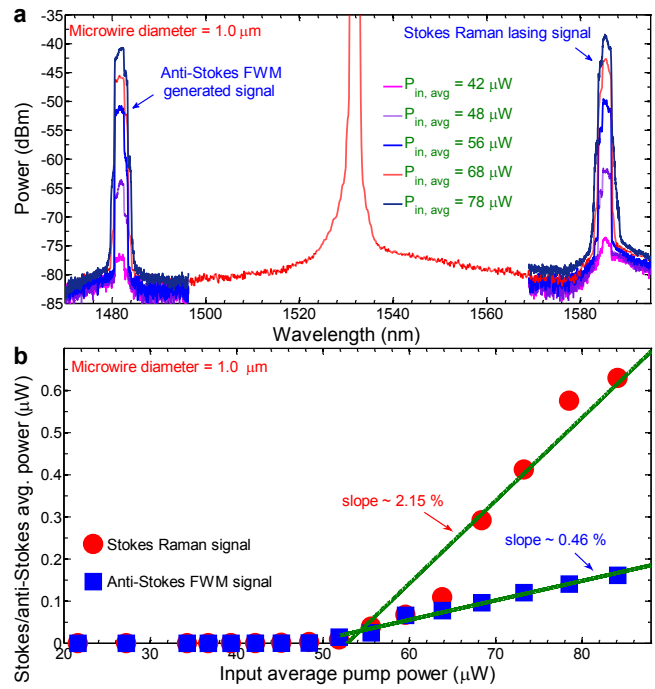

**FIG. 3.** (a) Output spectra showing the simultaneous operation of microwire Raman laser and FWM wavelength converter at different input pump power levels (b) Stokes and anti-Stokes slope efficiency curves.

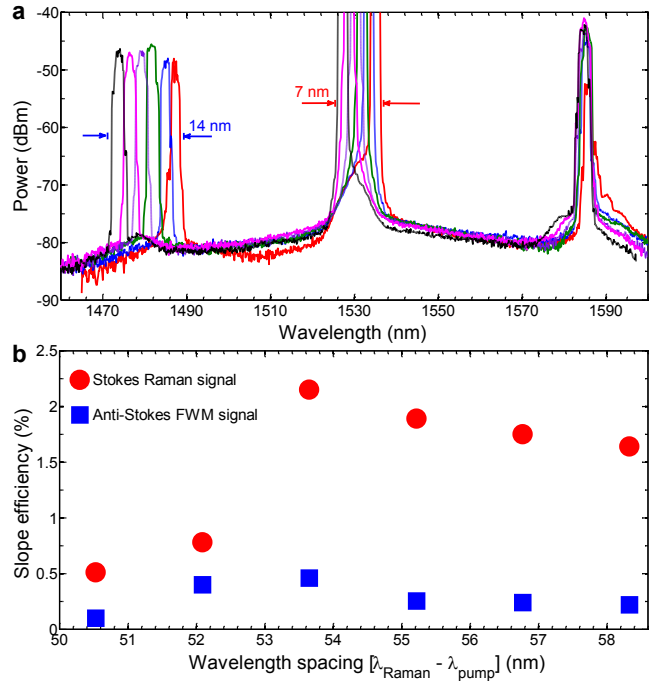

**FIG. 4.** (a) Spectra of a microwire Raman laser-FWM wavelength converter at different pump wavelengths (b) Slope efficiencies of (Stokes) Raman signal and (anti-Stokes) FWM signals for varying wavelength separation between the input pump and the resulting Raman laser.

18